\documentclass[12pt]{article}
\usepackage[]{graphicx}\usepackage[]{color}
\makeatletter
\def\maxwidth{ %
	\ifdim\Gin@nat@width>\linewidth
	\linewidth
	\else
	\Gin@nat@width
	\fi
}
\makeatother

\definecolor{fgcolor}{rgb}{0.345, 0.345, 0.345}

\usepackage{framed}
\makeatletter
{\par\unskip\endMakeFramed%
	\at@end@of@kframe}
\makeatother

\definecolor{shadecolor}{rgb}{.97, .97, .97}
\definecolor{messagecolor}{rgb}{0, 0, 0}
\definecolor{warningcolor}{rgb}{1, 0, 1}
\definecolor{errorcolor}{rgb}{1, 0, 0}
\newenvironment{knitrout}{}{} 

\usepackage{alltt}

\usepackage{mathptmx}       
\usepackage{helvet}         
\usepackage{courier}        
\usepackage{type1cm}        
\usepackage{bbm}
\usepackage{graphicx}        
\usepackage{multicol}        
\usepackage{amsmath}
\usepackage{amsfonts}
\usepackage{natbib}

\newcommand{\R}{\mathbb{R}}

\newcommand{\lp}{\left(}
\newcommand{\rp}{\right)}

\providecommand{\abs}[1]{\left|#1\right|}

\newcommand{\var}[1]{\textrm{Var}\lp#1\rp}
\newcommand{\lrp}[1]{\left(#1\right)}

\newcommand{\lrb}[1]{\left\{#1\right\}}

\newcommand{\norm}[1]{\left|\left| #1\right|\right|}

\newcommand{\inprod}[2]{\left<#1,#2\right>}
\newcommand{\mcal}[1]{\mathcal{#1}}
\newcommand{\Ind}[1]{\mathbbm{1}_{\lrb{#1}}}

\IfFileExists{upquote.sty}{\usepackage{upquote}}{}

\begin{document}

\title{Variable selection in Functional Additive Regression Models}

\author{Manuel Febrero--Bande\textsuperscript{1,2}, Wenceslao Gonz\'alez--Manteiga\textsuperscript{1,2}\\ and Manuel Oviedo de la Fuente\textsuperscript{1,2}}
\footnotetext[1]{ MODESTYA Group, Department of Statistics, Mathematical Analysis and Optimization, Universidade de Santiago de Compostela, Campus Vida, Santiago de Compostela, Spain.}
\footnotetext[2]{	 Technological Institute for Industrial Mathematics (ITMATI), Campus Vida, Santiago de Compostela, Spain.}


\maketitle

\abstract{This paper considers the problem of variable selection in regression models in the case of functional variables that may be mixed with other type of variables (scalar, multivariate, directional, etc.). Our proposal begins with a simple null model and sequentially selects a new variable to be incorporated into the model based on the use of distance correlation proposed by \cite{Szekely2007}. For the sake of simplicity, this paper only uses additive models. However, the proposed algorithm may assess the type of contribution (linear, non linear, ...) of each variable. The algorithm has shown quite promising results when applied to simulations and real data sets.}

\section{Introduction}
\label{sec:Introduction}
The variable selection problem in a general regression model tries to find the subset of covariates that best predicts or explains a response. In the classical approach, the covariates and the response are scalar (or multivariate) and the model established among them is linear.  

The stepwise regression, the most widely-used model selection technique throughout the 80's and the 90's, is rooted in the classical papers by \cite{Akaike1973}, \cite{Mallows1973}, \cite{Schwarz1978} and \cite{Stone1979}. The main idea is to use some diagnostic tools, directly derived from the linear model, to evaluate the contribution of a new covariate  and decide whether it should be included in the model. The final subset is usually constructed using two main strategies: 
the forward selection that begins with a simple null model and tests, at each step, the inclusion of a new covariate in the model; and the backward selection that starts with the full model including all candidate variables and removes the most insignificant one at each step. It is also possible to mix both strategies, testing at each step which variables can be included or excluded in the optimal regression subset of covariates. In any case, the stepwise regression is anchored in the diagnostics of the linear model; it is therefore blind to detecting contributions other than the linear one among the covariates and the response. 

The work by \cite{Tibshirani1996} proposing the LASSO estimator opens a new direction on variable selection procedures. The main innovation of the LASSO estimator is that it includes a $l_1$-type constraint for the coefficient vector $\beta$ to force some parameters (components of $\beta$) to equal zero and, thereby, obtains the optimal subset of covariates such as those with non-zero coefficients. The effect of the constraint also helps in the optimization step and satisfactorily deals with the sparsity phenomenon. See \cite{Zou2006} for a revision of the oracle properties of LASSO.
Interesting examples may be found in the literature following the same line but using several penalties, constraints or using different structures in the regression models, such as: LARS (\cite{Efron2002}), SCAD (\cite{Fan2001}), COSSO (\cite{Lin2006}), additive models (\cite{Xue2009}) and extensions to partial linear, additive or semiparametric models like PLM (\cite{Du2012}), APLM (\cite{Liu2011}) or GAPLM (\cite{Wang2011}). All these works have two common characteristics: each paper is based on a specific model and all the covariates must be included in the model at the same time. The latter could lead to highly demanding computing algorithms that are difficult to implement at times. In particular, for high-dimensional or functional data problems, the previous steps that commonly include variable standardization and/or variable representation may notably increase the complexity and the cost of the algorithms. See, for example, \cite{Hastie2015} for a review of some of the aforementioned methods.

Another type of strategy is a pure feature selection where the covariate is selected without a model. This is the approach employed in mRMR (minimum Redundancy Maximum Relevance, \cite{Peng2005}). To enter into the model, a new candidate covariate must have a great relevancy with the response while maintaining a lower redundancy with the covariates already selected in the model. The main advantage of this approach is that it is an incremental rule; once a variate has been selected, it cannot be deselected in a later step. On the other hand, the measures for redundancy and relevancy must be chosen in function of the regression model applied to ensure good predictive results in the final model.
The FLASH method proposed in \cite{Radchenko2011} is a modification of the LASSO technique that sequentially includes a new variate changing the penalty at each step. 
This greedy increasing strategy is also employed by Boosting (see, for example, \cite{Buhlmann2003} or \cite{Ferraty2009} in a functional data context). Boosting is not a purely feature selection method but rather a predictive procedure that selects at each step the best covariate/model with respect to the unexplained part of the response. The final prediction is constructed as a combination of the different steps. 
All the previous solutions are not completely satisfactory in a functional data framework, specially when the number of possible covariates can be arbitrarily large. Specifically, we are interested in an automatic regression procedure capable of dealing with a large number of covariates of different nature, possibly very closely related to one another.

\begin{knitrout}\footnotesize
	\definecolor{shadecolor}{rgb}{0.969, 0.969, 0.969}\color{fgcolor}\begin{figure}[!b]
		
		{\centering \includegraphics[width=.9\linewidth]{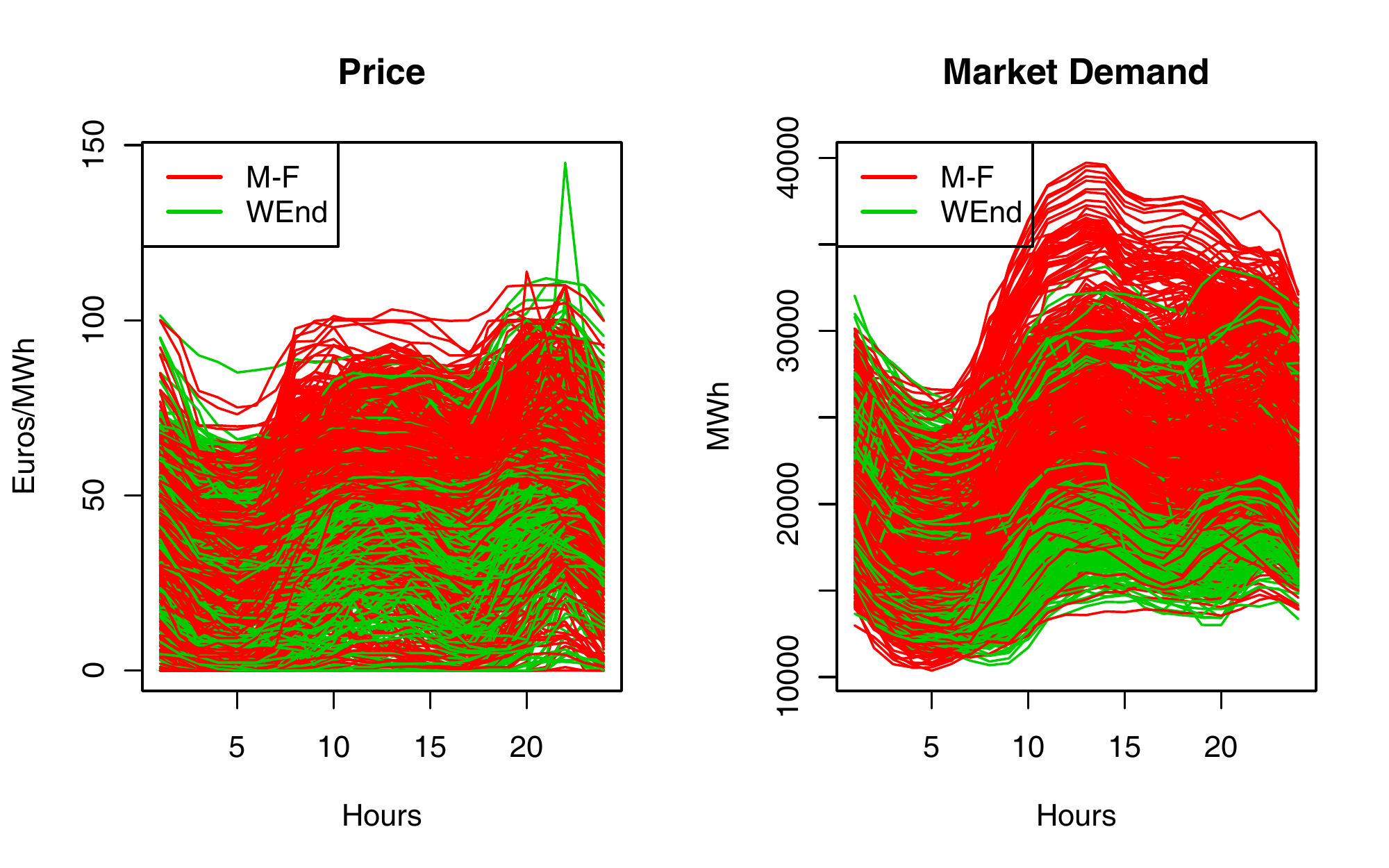} 
			
		}
		
		\caption[Price and Energy Market Demand curves negotiated in the Iberian Energy Market]{Price and Energy Market Demand curves negotiated in the Iberian Energy Market. Period: Jan, 2014- Dec, 2014 (Source: omie.es).}\label{fig:energy}
	\end{figure}	
\end{knitrout}

\begin{figure}[htb]
	\footnotesize		
		{\centering
			\includegraphics[width=.9\linewidth]{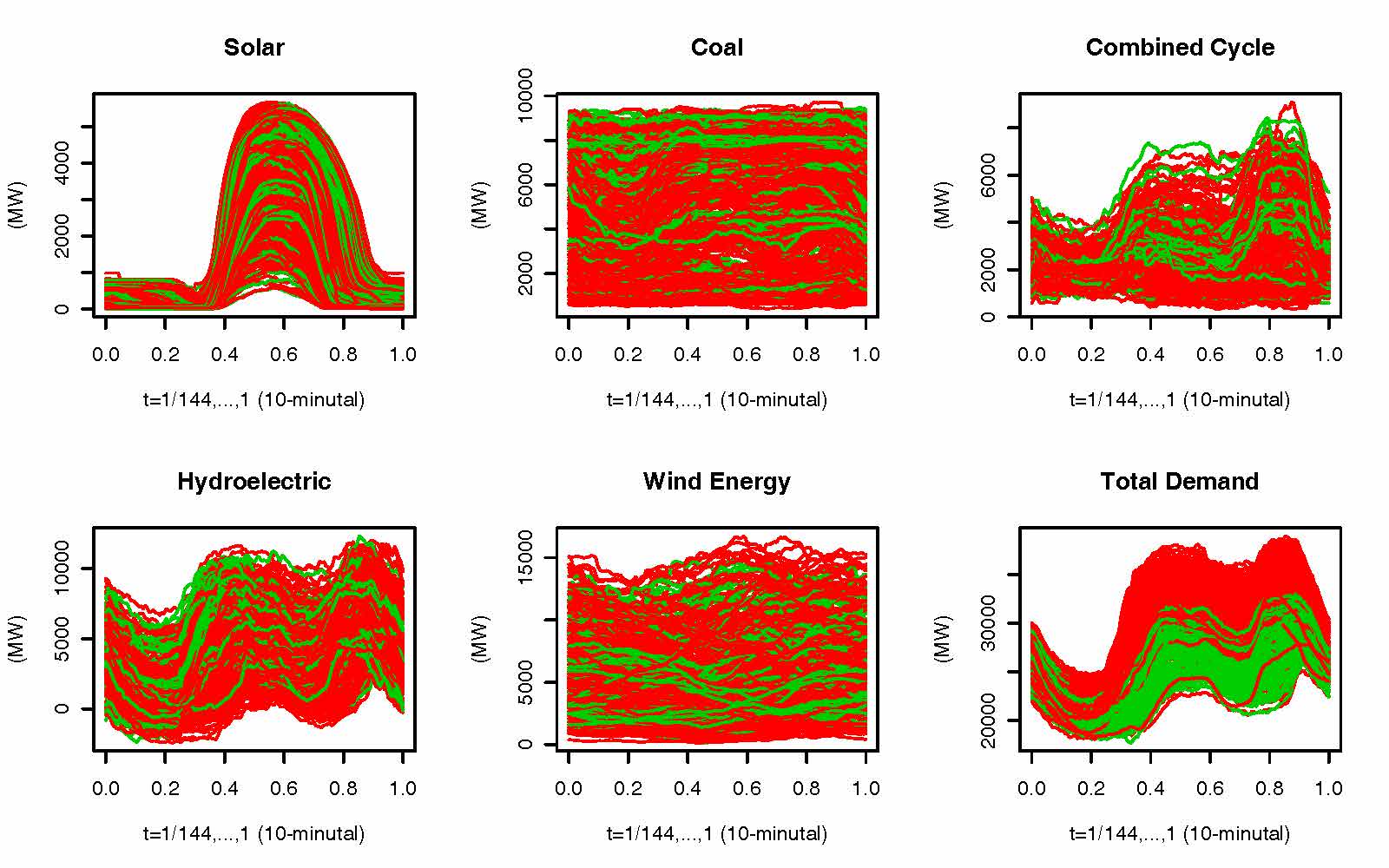}	
		}
		
		\caption[Daily profile curves of amount of generated energy by type]{Daily profile curves of amount of generated energy by type. Period: Jan, 2014- Dec, 2014 (Source: demanda.ree.es). Red and green colors represents, respectively, M-F and W-End days.}\label{fig:energy2}
	\end{figure}

Our motivating example comes from the energy market. Figure~\ref{fig:energy} shows the daily profile of Price and Energy (Electricity) Market Demand (both measured hourly) at the Iberian Energy Market from Jan, 2014 to Dec, 2014 with a color code, red and green, that considers, respectively, two groups: M-F (Monday to Friday) and WEnd (Saturday and Sunday \& Sun). We are interested in predicting the price or the demand at certain hours of the following day. We dispose of many different sources of information to do this. These sources may include other variables related with energy market or generation, meteorological information, calendar effects or any transformation/filter of the preceding. Figures \ref{fig:energy2} and \ref{fig:energy3} show a small sample of selected covariates. Figure~\ref{fig:energy2} includes some of the energy generation variates (by type) included in the energy generation pool in the same period. The total demand can be decomposed as a sum of the different energy generations covering the demand. Figure~\ref{fig:energy3} is a small sample of calendar and meteorological information. In both figures, the color code is the same as in Figure~\ref{fig:energy}. Note that the number of variables that may be included in the model is rather high (over $150$).

The curves in Figure~\ref{fig:energy2} are the components of the energy pool by generation type, which may be useful for predicting price or demand. Due to the Iberian Energy Market regulations, the odds of becoming part of the final energy pool consumed due to its price or availability are unalike among all the types of energy. For instance, hydroelectric energy is only offered to the market when the price is high in the presence of founded expectations on refilling the reservoir (using the weather forecasts).

\begin{knitrout}\footnotesize
	\definecolor{shadecolor}{rgb}{0.969, 0.969, 0.969}\color{fgcolor}\begin{figure}[!t]		
		{\centering \includegraphics[width=.9\linewidth]{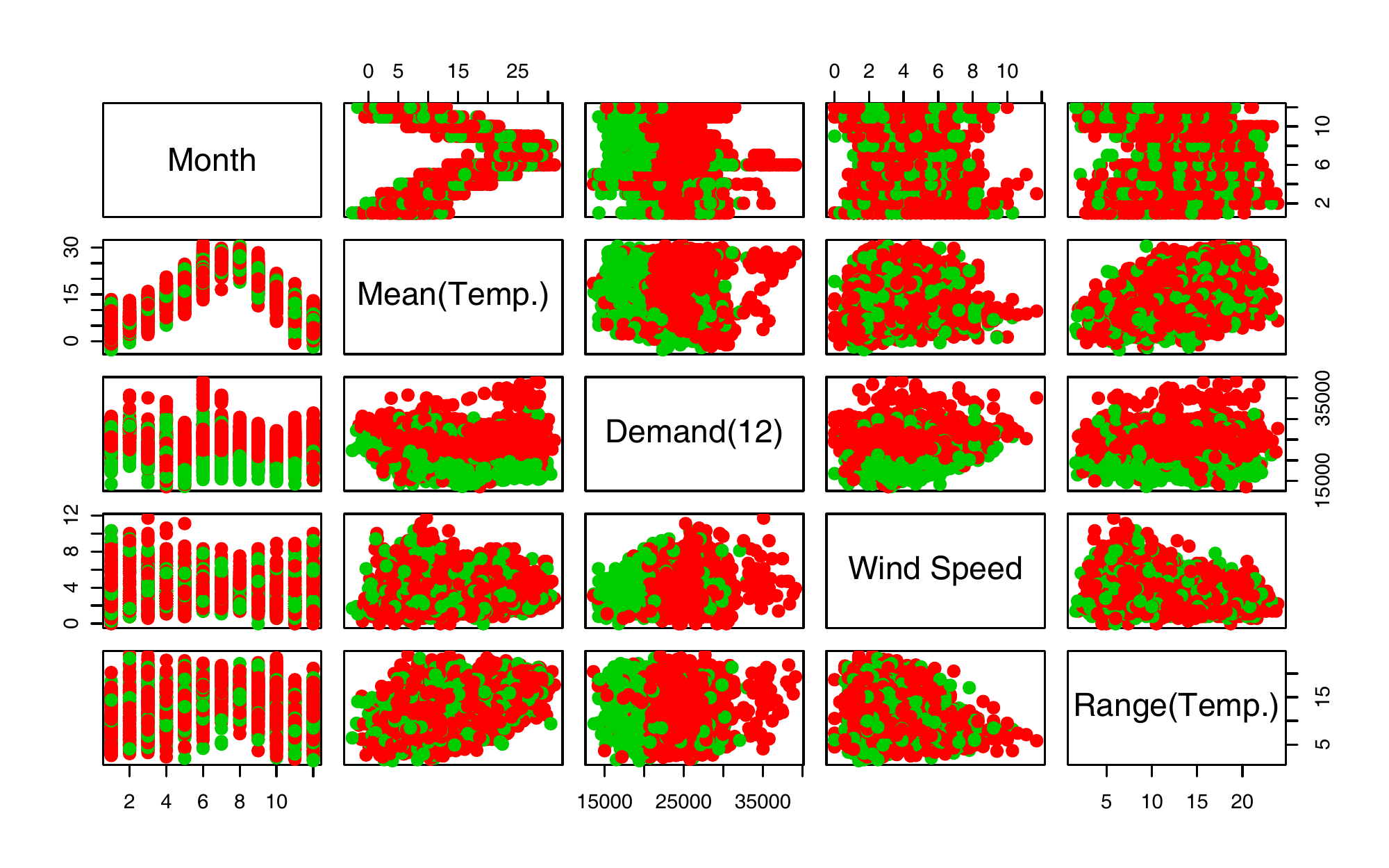} 			
		}		
		\caption[Examples of available scalar covariates for the energy prediction problem]{Examples of available scalar covariates for the energy prediction problem. Period: Jan, 2014- Dec, 2014 (Source: aemet.es). Red and green colors represents, respectively, M-F and W-End days.}\label{fig:energy3}
	\end{figure}	
\end{knitrout}

The scalar covariates for our prediction problem can be linked to calendar effects (like month or day-of-week), meteorological information (like temperature or wind speed) and transformations from the functional variables (such as the market demand at midday $X(t)\Ind{t=12h}$). This means that many new variables can be created from the original ones (for instance, by using derivatives or considering certain subintervals), many of which may certainly be closely related.

Our aim is to select significant covariates for a general additive regression model with scalar response:    
\begin{equation}\label{eq:AM}
	Y_i=\alpha+\sum_{j=1}^Jf_j\lrp{X_i^{(j)}}+\varepsilon_i,\quad i=1,\ldots,N
\end{equation}
where the covariates are chosen from the set $S=\lrb{X^1,X^2,\cdots,X^k,\cdots}$ of different potential covariates (functional, vectorial, \ldots). The notation $X^{(j)}$ refers to the $j$-th covariate selected for the model. The number of variates can be extraordinarily large, so we intent to construct the regression model sequentially, i.e. from the trivial model up to the one that includes all the useful information provided by the covariates in the set.

The rest of this paper is structured as follows. Section~\ref{sec:Alg} presents our proposal, Section~\ref{sec:Numresults} presents some artificial examples and simulation studies and 
Section~\ref{sec:appl} includes the applications to our motivating example.  

\section{The procedure}
\label{sec:Alg}
Our proposal borrows some ideas from other variable selection methods. First, we separate the selection process from the estimation step by picking the possible candidates using a measure that is not based in a concrete model. The method for selecting the covariate must be general so that the model may include scalar, multivariate, directional or functional covariates in the model on an impartial way. A model is chosen (among those in a catalog) that can be applicable to the selected covariates and the response, ensuring that the model is correct, i.e. it fulfills the considered structural hypothesis and its performance is superior to the previous one. Typically, we must decide the contribution of the new variable jointly with the selected ones in precedent steps. The new model is compared with the previous one using an appropriate test to decide whether to definitely incorporate the new variate into the model. If the new model is not better than the previous one, the candidate is discarded and the current model remains invariant. If the new model is better then the new model becomes the current model. Finally, using the residuals of the current model, new candidates are trying following the same steps. This procedure is repeated until no more variables enter the model. The idea of using residuals makes that, in subsequent steps, the algorithm attempts to capture information not chosen by previous steps. This also sharply contrasts with the rules of a pure feature method. In a pure feature method it is necessary to compute a measure of redundancy among the covariates in the model and the candidates to be incorporated that can be quite difficult due to different nature of the variables.

As aforementioned, the key idea for the selection of a feature is to find a tool that can be homogeneously applied to variates of different nature. This paper is based on the exhaustive use of the distance correlation, $\mathcal{R}$, proposed by \cite{Szekely2007}, and recently reviewed in \cite{Szekely2017}. The distance correlation fulfills the following two conditions:
\begin{itemize}
	\item[i)] $\mathcal{R}(X,Y)$ is defined for $X$ and $Y$ random vector variables in arbitrary finite dimension spaces.
	\item[ii)] $\mathcal{R}(X,Y)=0$ characterizes independence of $X$ and $Y$.
\end{itemize}

Both conditions that are established in \cite{Szekely2007}, signal the presence of a way to measure the relationship among $X$ and $Y$ homogeneously for arbitrary dimensions of these vectors. Indeed, the work by \cite{Lyons2013} extends these properties for metric spaces of a certain type (\emph{strong negative type}). In particular, Hilbert spaces or any space that can be embedded in a Hilbert space are of strong negative type. Along this paper, we will only use Hilbert spaces for all covariates (for scalar covariates this means that we are using the classical Euclidean spaces). As $\mcal{R}$ characterizes independence, the distance correlation can detect relationships among variables other than the linear one. A test for independence using the distance correlation can be derived following  \cite{Szekely2007} and \cite{Szekely2013} that allows us to contrast when a variable could be a reasonable candidate to enter the model. 

In \cite{Yenigun2015}, a definition of partial distance correlation (PDC) among $X$ and $Y$ given $Z$ was introduced based on computing the distance correlation among the residuals of two models: $Y$ respect to $Z$ and $X$ respect to $Z$. The PDC is the key of that paper for a variable selection method although has two important drawbacks: First, PDC is constructed under linear relationship assumptions among variables and, in the original paper, it is only applied to classical linear models with scalar variates (the authors called it ``linear PDC''). Second, its implementation is done using the function \texttt{pdcor} of \texttt{R}--package \texttt{energy} (\cite{Rizzo2014}) which only uses the distance matrices among elements of $X$, $Y$ and $Z$. Specifically, $Z$ (the variables already in the model) could be a mix of functional, scalar or multivariate variables where an appropriate distance using all of them must be hard to compute. Even restricting ourselves to the scalar case, those variables should have a similar scale. On the contrary, our proposal can be applied to any mix of variables given that in every step, the distance correlation is computed among the residuals of the current model with each candidate. Taking into account that the residuals have the same nature as the response variable, the distance correlation can always be computed at each step, supposing that we were able to compute $\mcal{R}(X,Y)$ at first iteration.

The computation of $\mcal{R}(X,Y)$ for a given sample (empirical distance correlation) is quite straightforward because it only depends on the distances among data. See again, \cite{Szekely2007} for details. Specifically, supposing we are interested in computing the distance correlation among $X^j$ and $Y$, let $a_{kl}=d(X_k^{j},X_l^{j})$, the distance in the Hilbert space of $X^{j}$  among cases $k$ and $l$, and $A_{kl}=a_{kl}-\bar{a}_{k\cdot}-\bar{a}_{\cdot l}+\bar{a}_{\cdot\cdot}$ and respectively, 
$b_{kl}=d(Y_k,Y_l)$, and $B_{kl}=b_{kl}-\bar{b}_{k\cdot}-\bar{b}_{\cdot l}+\bar{b}_{\cdot\cdot}$.

The distance correlation is simply computed as:
\[
\mathcal{R}(X^j,Y)=\frac{\sum_k\sum_{l} A_{kl}B_{kl}}{\sqrt{\sum_k\sum_{l} A_{kl}^2}\sqrt{\sum_k\sum_{l} B_{kl}^2}}
\]

The simplicity of the $\mcal{R}(X^j,Y)$ has an important drawback in terms of computational time. As it is based on distances, the memory and the number of necessary operations is of order $N(N-1)/2$ where $N$ is the sample size. So, for large sample sizes, the computation must be divide into some small tasks to avoid memory overruns. See the end of section~\ref{sec:appl} for details on how to solve this issue.

\subsection{The algorithm}

Our proposal can be formalized as follows: 

\begin{enumerate}
	\item Let $Y$ the response and $S=\lrb{X^1,\ldots,X^p}$ the set of all variables that can be included in the model. 
	\item Set $\hat{Y}=\bar{Y}$, and let $M^{(0)}=\emptyset$ the initial set of the variates included in the model. Set $i=0$.
	\item Compute the residuals of the current model: $\hat{\varepsilon}=Y-\hat{Y}$.
	\item Choose $X^j\in S\ne\emptyset$ such that: 1) $\mathcal{R}\lrp{\hat{\varepsilon},X^j}\ge \mathcal{R}\lrp{\hat{\varepsilon},X^k}, \forall k\ne j\in S$ and 2) the null hypothesis for the test of independence among $\lrb{X^j}$ and $\hat{\varepsilon}$ is rejected. IF NOT, END.
	\item Update the sets $M$ and $S$: $M^{(i+1)}=M^{(i)}\cup\lrb{X^j}$, and $S=S\backslash\lrb{X^j}$.
	\item Compute the new model for $Y$ using $M^{(i+1)}$ choosing the best contribution of the new covariate. Typically, there will be a catalog of all possible ways of constructing correct models with the variates in $M^{(i+1)}$ fixing the contributions of the variates in $M^{(i)}$ and adding the new one. 
	\item Analyze the contribution of $X^j$ in the new model respect to the current: \newline IF this contribution is not relevant (typically comparing with the current model)\newline THEN $M^{(i+1)}=M^{(i+1)}\backslash\lrb{X^j}$ and the current model remains unalterable \newline ELSE the new model becomes the current model and provides new predictions ($\hat{Y}$). Along this paper we have employed an additive model: $\hat{Y}=\bar{Y}+\sum_{m\in M}\hat{f}_m\lrp{X^{(m)}}$ where at each step $\hat{f}_m$ could be linear or nonlinear.
	\item Update the number of iterations: $i=i+1$ and go to 3
	\item END. The current model is the final model with the variates included in $M^{(i)}$. $S$ is either the empty set or contains those variables that accept the null hypothesis of the test of independence respect to the residuals of the current model. 
\end{enumerate}

Steps 1--3 establish the null model as the initial model for beginning the procedure. Step 4 selects the variable from the set of available ones $S$. A new candidate is selected when maximizes the distance correlation among available ones and the test of independence between $\hat{\varepsilon}$ and the candidate rejects the null hypothesis. For simplicity, every time a distance correlation is showed, the value is filtered by the test of independence, i.e. all tables show $\mathcal{R}\lrp{X^k,\hat{\varepsilon}}\Ind{H_1}$ that only takes values distinct from zero for those covariates that rejects the null hypothesis of independence. From the practical point of view, the algorithm spends most of its time computing the latter quantity for each variable in this step. This task is not so fast because the distance matrices for a sample size $N$ require $N(N-1)/2-N$ stored positions; so, both the number of operations and the amount of consumed memory are quadratic functions of the sample size. We may alleviate the computational burden using two simple tricks: first, we may easily parallelize this task to into small tasks that compute the distance correlation separately for different covariates in $S$ or even, when the distance matrix is too large, the computation of $\mathcal{R}$ can be done through all the submatrices below the diagonal. Secondly, we can compute the distance matrices needed for the algorithm in advance for all covariates avoiding a repetitive calculation of these matrices for each iteration.

Step 5 updates the sets of variates. The selected variable is included as a possible candidate in set $M$ and it is removed from set $S$. So, each candidate is only checked once in a clear forward strategy. 

Step 6 tries the possible models that can be conformed with the variables included in the updated $M$. Here, the previous contributions of the variables already in the model remain fixed and the catalog is used to check what could be the contribution of the new candidate. For instance, it may happen that my catalog of models only contains linear models or models with a limited number of nonlinear terms. In the latter case, if that limit is reached in previous steps, the new candidate can only be included with a linear contribution. This means that the catalog of possible models may not suffice to explain the relationship shown by the distance correlation. As a diagnostic tool, in latter iterations the test of independence can be computed for those variates in $M$ to analyze if the contribution of covariate already in the model is collecting all the possible information of that covariate. If the test is rejected for any of the variables of $M$, it is an evidence that the model is not completely exploiting its information. Please note that this diagnostic tool is never done in step 5 to avoid an infinite loop. Of course, a final arrangement is leaved to the user when the procedure ends. For instance, a certain variate entering the model at an earlier stage, may become irrelevant some iterations later, when other variates add their contributions to the model. Of course, a final diagnostic of the model could lead to changes in the contribution of the covariates. Yet the comparison among the different variables must be carefully analyzed given its nature. Of course, the procedure could be modified to include backward steps as in the classical stepwise method but, also due the distinct nature of the covariates, it could be difficult to find an homogenize backward step that works in an universal way. To avoid this type of problems, we have only applied additive models in this paper because this type of model is available for (nearly) any kind of variates and is quite flexible taking into account that each contribution may be linear or nonlinear (smooth). For instance, if we want to include iterations among variables, we can define a new covariate computing that interaction to be included as a candidate in the additive model.

Step 7 analyzes the contribution of the new candidate in the model with respect to the previous one. Typically, depending on the model, this can be done with a Generalized Likelihood Ratio Test. If the contribution of the new candidate is not relevant, the variate is removed from set $M$ and the previous model remains as the current model trying with another candidates. If the new model is better than the previous one, the new residuals are computed and the algorithm backs to step 3 to recompute the distance correlation among these residuals and the variables remaining in set $S$. The procedure ends when no more variables can be added to the model because $S$ is empty or all variables in $S$ accepts the independence null hypothesis.     

\section{Numerical results}
\label{sec:Numresults}
In the general framework, there is no previous papers that can be considered direct competitors of our proposal. But, as commented before, \cite{Yenigun2015} proposes an algorithm valid for scalar variates under linear relationships with the response. We have compared our proposal in the same scenarios of that paper. As in the original paper,  samples of size $N=100$ are generated from different regression models and with several distributions for the $p=8$ candidate covariates (typically combinations of $N(0,\sigma^2)$ and $U[-a,a]$). This process was repeated $B=500$ times and the number of times every single covariate was selected is recorded as well as the Root Mean Squared Prediction Error: $$RMSPE=\frac{1}{B}\sum_{i=1}^{B}\sqrt{\frac{1}{n}\sum_{j=1}^{n}(Y_j^{(i)}-\hat{Y}_j^{(i)})^2}$$ for the final model using a test sample also of size $n=100$. The scenarios are defined as follows:
\begin{enumerate}
	\item[YR1:] $Y=Z_1+Z_2+Z_3+\varepsilon,\quad Z_i\sim N(0,1), \; \varepsilon\sim N(0,\sigma^2=2^2)$
	\item[YR2:] $Y=\log[4+\sin(3Z_1)+\sin(Z_2)+Z_3^2+Z_4+0.1\varepsilon], \quad Z_1\sim N(0,1), \; Z_2\sim N(0,2^2), \\ Z_3\sim U[-1.5,1.5], \; Z_4,\ldots,Z_8\sim U[-1,1]$ and $\varepsilon\sim N(0,1)$
	\item[YR3:] $Y=\abs{Z_1}+Z_2^2+Z_3^2, \quad Z_1\sim N(0,1.4^2), \; Z_2\sim U[-1.7,1.7], \; Z_3\sim N(0,.8^2), \\ Z_4,\ldots,Z_8\sim N(0,1)$.
	\item[YR4:] $Y=Z_1+Z_2+Z_3+\varepsilon, \; Z\sim N_p(0,\Sigma)$ with $\Sigma=\lrp{\begin{array}{cccc} 1 & \theta & \cdots & \theta \\ \theta & 1 & \cdots & \theta \\ \vdots & \vdots & \ddots & \vdots \\ \theta & \theta & \cdots & 1\end{array}}$ where $\theta=0.6$ and $\varepsilon\sim N(0,2^2)$.
	\item[YR5:] $Y=Z_1+Z_2+Z_3+\varepsilon, \;\\ Z\sim N_p(0,\Sigma)$ with $\Sigma=\lrp{\begin{array}{ccccc} 1 & \theta & \theta^2 & \cdots & \theta^{p-1} \\ \theta & 1 & \theta & \cdots & \theta^{p-2} \\ \theta^2 & \theta & 1 & \cdots & \theta^{p-3}\\ \vdots & \vdots & \vdots & \ddots & \vdots \\ \theta^{p-1} & \theta^{p-2} & \theta^{p-3} & \cdots & 1\end{array}}$ \\ where $\theta=0.6$ and $\varepsilon\sim N(0,2^2)$.
\end{enumerate}  

For these scenarios, we have compared the proposal of \cite{Yenigun2015} (PDC) with ours in two versions: restricting the model to be have only linear contributions (LM) or letting every contribution to be nonlinear (AM). The results are provided in Table~\ref{tab:YR}. In the independent linear case YR1, the three methods show a similar performance both in variable identification and in terms of prediction error. In the second scenario, YR2, the constraint to deal only with linear models produces high values of RMSPE for PDC and LM on the contrary of the AM case that mimics the theoretical $\sigma$. Indeed, $Z_1$ is not identified the same proportion of times than $Z_2, Z_3$ or $Z_4$ except in the case of AM. The third scenario, YR3, shows similar results than YR2 although here the difference among linear versions and the nonlinear one is now more important in terms of RMSPE. Recall that these two scenarios are nonlinear. Scenarios YR4 and YR5 are linear scenarios with dependence among covariates. In both cases, it seems that our proposals (LM or AM) identify the relevant covariates less accuracy than PDC but really, PDC tends to include more variates than necessary. In YR4, the minimum of proportion of times that $Z_4,\ldots, Z_8$ enters the model is $0.436$ where for LM and AM the maximum that a irrelevant covariate enters the model is $0.044$. Something similar happens in YR5 although due to the dependence structure, only $Z_4$ enters the model a high proportion of times with the PDC. The difference is that YR4 considers a constant correlation among all covariates but in YR5, only $Z_4$ shows an important correlation with $Z_3$ (the correlation decreases respect to the column number). Taking into account the RSMPE, it seems that LM and AM do not need to select all the relevant variables to achieve similar prediction levels. On the contrary, PDC needs to include some of the irrelevant ones as part of the model. Intuitively, a high dependence among covariates leads to that not all the relevant variates may enter the model because, and due to the dependence structure, its contribution may have been incorporated into the model by other highly related. So, as a conclusion of this small simulation study, our general proposal AM works well in linear scenarios, clearly better in nonlinear ones and in the case of dependence among covariates reduces the amount of irrelevant variates that are introduced in the model without losing predictive performance.   

\begin{table}[ht]
	\centering
	\caption{ Proportion of times that each variable enters the model for scenarios YR1--YR5. The last column contains the Root Mean Square Prediction Error for $500$ repetitions.} 
	\label{tab:YR}
	\footnotesize
	\begin{tabular}{|c|c|c|c|c|c|c|c|c||c|}
		\hline
		& $Z_{1}$ & $Z_{2}$ & $Z_{3}$ & $Z_{4}$ & $Z_{5}$ & $Z_{6}$ & $Z_{7}$ & $Z_{8}$ & RMSPE \\ 
		\hline
		YR1--PDC & 0.968 & 0.968 & 0.976 & 0.078 & 0.096 & 0.082 & 0.078 & 0.080 & 2.070 \\ 
		YR1--LM & 0.990 & 0.996 & 0.998 & 0.050 & 0.096 & 0.076 & 0.054 & 0.086 & 2.060 \\ 
		YR1--AM & 0.986 & 0.994 & 0.994 & 0.046 & 0.084 & 0.066 & 0.056 & 0.078 & 2.110 \\ 
		\hline
		YR2--PDC & 0.658 & 0.980 & 0.886 & 0.996 & 0.066 & 0.060 & 0.082 & 0.080 & 0.280 \\ 
		YR2--LM & 0.776 & 0.996 & 0.952 & 0.994 & 0.050 & 0.064 & 0.068 & 0.062 & 0.280 \\ 
		YR2--AM & 1.000 & 1.000 & 1.000 & 1.000 & 0.052 & 0.054 & 0.078 & 0.066 & 0.090 \\ 
		\hline
		YR3--PDC & 1.000 & 0.996 & 0.948 & 0.068 & 0.080 & 0.064 & 0.084 & 0.072 & 1.560 \\ 
		YR3--LM & 0.994 & 0.994 & 0.934 & 0.074 & 0.064 & 0.060 & 0.076 & 0.068 & 1.560 \\ 
		YR3--AM & 1.000 & 1.000 & 1.000 & 0.076 & 0.054 & 0.056 & 0.048 & 0.078 & 0.060 \\ 
		\hline
		YR4--PDC & 0.992 & 0.988 & 0.998 & 0.436 & 0.476 & 0.452 & 0.464 & 0.468 & 2.060 \\ 
		YR4--LM & 0.766 & 0.792 & 0.772 & 0.044 & 0.040 & 0.040 & 0.022 & 0.034 & 2.130 \\ 
		YR4--AM & 0.754 & 0.776 & 0.764 & 0.038 & 0.038 & 0.038 & 0.014 & 0.032 & 2.180 \\ 
		\hline
		YR5--PDC & 0.998 & 1.000 & 0.998 & 0.350 & 0.042 & 0.014 & 0.020 & 0.016 & 2.040 \\ 
		YR5--LM & 0.876 & 0.856 & 0.820 & 0.082 & 0.058 & 0.034 & 0.046 & 0.050 & 2.100 \\ 
		YR5--AM & 0.876 & 0.862 & 0.808 & 0.086 & 0.060 & 0.044 & 0.038 & 0.062 & 2.150 \\ 
		\hline
	\end{tabular}
\end{table}

For assessing the situation when the variables are of different nature, we have developed a simulation study to check the performance of the algorithm in a mixed scenario with functional and scalar variables. Five functional and five scalar variables were simulated and the response was constructed as a function of the two first functional and the two first scalar variables. The functional variables: $\lrb{\mcal{X}_1,\ldots,\mcal{X}_5}$ are generated following independent Ornstein-Uhlenbeck processes in $[0,1]$, and the scalar variables $\lrb{Z_1,\ldots,Z_5}$ following, respectively,  $U[0,1]$, $N(0,1)$, $N(0,1)$, $U[0,1]$ and $N(0,1)$ independently of each other.
We constructed the response as follows: 
\[
Y=10+a_1\inprod{\mcal{X}_1}{\beta_1}+a_2\norm{\mcal{X}_2^2}+3a_3Z_1+a_4Z_2^2+\varepsilon
\]
with $\beta_1=2t+\sin{4\pi t+0.1}, t\in [0,1]$ and $\varepsilon\sim N(0,.25^2)$.

The coefficients $a=\lrb{a_1,a_2,a_3,a_4}$ were introduced to emphasize/mask each part of the model. We estimated the model through a Functional Additive Model (\cite{Muller2008}) using the first four principal components for functional covariates and a standard additive model for the scalar ones. Samples with $N=200$  were generated and the process was repeated $B=500$ times to count the proportion that a particular covariate enters the model. Table~\ref{tab:sim} shows the excellent results for different combinations of $\lrb{a_1,a_2,a_3,a_4}$. In all cases, $\mcal{X}_1$, $Z_1$ and $Z_2$ were selected all the times in the $500$ repetitions. $\mcal{X}_2$ was also selected all the times except for $a_2=\frac{1}{8}$ that corresponds to a small contribution of that variate to the model. This can be explained because the $\var{0.125\norm{\mcal{X}_2^2}}\approx 0.01$ that it is about $\frac{1}{6}$ of the residual variance. The non relevant variables enter the model about 5\% of the times. 

\begin{table}[h]
	
	\caption{\% of times that the variate was included in the model for $B=500$ replications. Independent case.\label{tab:sim}}
	\footnotesize
	\begin{tabular}{|c||c|c|c|c|c||c|c|c|c|c|}
		\hline
		$a$ & $\mcal{X}_1$ & $\mcal{X}_2$ & $\mcal{X}_3$ & $\mcal{X}_4$ & $\mcal{X}_5$& $Z_1$ & $Z_2$ & $Z_3$ & $Z_4$ & $Z_5$ \\ \hline
		$\lrb{1,1,1,1}$ & 1.000 & 1.000 & 0.046 & 0.038 & 0.030 & 1 & 1 & 0.030 & 0.040 & 0.054 \\
		$\lrb{\frac{1}{2},\frac{1}{2},\frac{1}{2},\frac{1}{2}}$ & 1.000 & 1.000 & 0.048 & 0.044 & 0.046 & 1 & 1 & 0.060 & 0.060 & 0.046 \\
		$\lrb{\frac{1}{4},\frac{1}{4},\frac{1}{4},\frac{1}{4}}$ & 1.000 & 0.984 & 0.040 & 0.044 & 0.058 & 1 & 1 & 0.058 & 0.040 & 0.060 \\
		$\lrb{\frac{1}{4},\frac{1}{4},1,1}$ & 1.000 & 0.988 & 0.056 & 0.030 & 0.052 & 1 & 1 & 0.058 & 0.044 & 0.052 \\
		$\lrb{1,1,\frac{1}{4},\frac{1}{4}}$ & 1.000 & 1.000 & 0.058 & 0.052 & 0.042 & 1 & 1 & 0.048 & 0.064 & 0.056 \\
		$\lrb{\frac{1}{8},\frac{1}{8},\frac{1}{8},\frac{1}{8}}$ & 0.908 & 0.474 & 0.048 & 0.050 & 0.044 & 1 & 1 & 0.058 & 0.060 & 0.060\\
		\hline
	\end{tabular}
\end{table}

We have repeated the simulation study but forcing that some irrelevant covariates ($\mcal{X}_3$, $\mcal{X}_4$, $\mcal{X}_5$, $Z_3$, $Z_4$, $Z_5$) may have a strong relationship with any of the important ones ($\mcal{X}_1$, $\mcal{X}_2$, $Z_1$, $Z_2$) to check how the procedure approaches covariates with strong collinearities.

To this end, we computed new covariates $\mcal{X}_3^*$, $\mcal{X}_4^*$, $Z_3^*$ and $Z_4^*$ in the following way: $\mcal{X}_3^*=0.95\mcal{X}_1+0.05\mcal{X}_3$, $\mcal{X}_4^*=0.95\mcal{X}_2+0.05\mcal{X}_4$,  $Z_3^*=0.95Z_1+0.05Z_3$ and $Z_4^*=0.95Z_2+0.05Z_4$. This ensures that $\mcal{X}_3^*$ (respectively, $\mcal{X}_4^*$, $Z_3^*$, $Z_4^*$) has nearly the same information as $\mcal{X}_1$ (respectively, $\mcal{X}_2$, $Z_1$, $Z_2$). 
Table~\ref{tab:sim2} shows the proportion of times that every variate is included in the model for this new scenario. Now the proportion of times shown in Table~\ref{tab:sim} for any of the relevant variables is shared among that variable and its (almost) copy. The important message is that when a variate is included in the model, its copy is banned for consequent steps and the model can be safely estimated without redundant information. Typically now, the sum of proportion of times of each covariate and its (almost) copy in Table~\ref{tab:sim2} is the proportion of times that can be seen in Table~\ref{tab:sim} for the covariate.

\begin{table}[h]
	\caption{\% of times that the variate was included in the model for $B=500$ replications under strong collinearities among covariates.\label{tab:sim2}}
	\footnotesize
	\begin{tabular}{|c||c|c|c|c|c||c|c|c|c|c|}
		\hline
		$a$ & $\mcal{X}_1$ & $\mcal{X}_2$ & $\mcal{X}_3^*$ & $\mcal{X}_4^*$ & $\mcal{X}_5$& $Z_1$ & $Z_2$ & $Z_3^*$ & $Z_4^*$ & $Z_5$ \\ \hline
		$\lrb{1,1,1,1}$ & 0.584 & 0.644 & 0.416 & 0.356 & 0.046 & 0.774 & 0.574 & 0.228 & 0.426 & 0.038 \\
		$\lrb{\frac{1}{2},\frac{1}{2},\frac{1}{2},\frac{1}{2}}$ & 0.550 & 0.640 & 0.450 & 0.360 & 0.048 & 0.738 & 0.580 & 0.262 & 0.420 & 0.052 \\
		$\lrb{\frac{1}{4},\frac{1}{4},\frac{1}{4},\frac{1}{4}}$ & 0.580 & 0.594 & 0.420 & 0.396 & 0.032 & 0.724 & 0.538 & 0.276 & 0.462 & 0.060 \\
		$\lrb{\frac{1}{4},\frac{1}{4},1,1}$ & 0.558 & 0.564 & 0.442 & 0.416 & 0.048 & 0.846 & 0.584 & 0.154 & 0.416 & 0.060 \\
		$\lrb{1,1,\frac{1}{4},\frac{1}{4}}$ & 0.592 & 0.684 & 0.408 & 0.316 & 0.040 & 0.662 & 0.540 & 0.338 & 0.460 & 0.072 \\
		$\lrb{\frac{1}{8},\frac{1}{8},\frac{1}{8},\frac{1}{8}}$ & 0.484 & 0.250 & 0.406 & 0.222 & 0.038 & 0.664 & 0.502 & 0.336 & 0.496 & 0.048\\
		\hline
	\end{tabular}
\end{table}


Table~\ref{tab:exampl} shows the distance correlation with the residuals for a single run at each iteration of the procedure under independence among covariates. The first column of this table is the iteration step. This allows us to analyze the order in which the covariates enter the model: $Z_1$ ($0.177$), $Z_2$ ($0.141$), $\mcal{X}_1$ ($0.22$), $\mcal{X}_2$ ($0.103$). In all cases, $\mathcal{R}(X,\hat{\varepsilon})$ increases with the iterations until that covariate is chosen. This is an expected behaviour in a case in which all the covariates are independent of each other and the relationship with the residual becomes stronger as previous effects of other covariates are removed. Also, if we compute $\mathcal{R}(X,\hat{\varepsilon})\Ind{\mathcal{H}_1}$ for $X \in M$, we obtain zeros suggesting that the model was able to incorporate all the important information of that covariates. In this case, the algorithm ends in iteration five when no more covariates are added to the model because none of the remaining ($\mcal{X}_3$, $\mcal{X}_4$, $\mcal{X}_5$, $Z_3$, $Z_4$, $Z_5$) rejects the null hypothesis of independence with respect to the residuals.

\begin{table}[ht]
	\centering
	\caption{ $\mathcal{R}(X,\hat{\varepsilon})\Ind{\mathcal{H}_1}$ for one run with $a=\lrb{1,1,1,1}$. At each iteration, a variable enters the model when maximizes the row. Empty cells correspond to variates selected in previous steps.} 
	\label{tab:exampl}
	\footnotesize
	\begin{tabular}{|c||ccccc||ccccc|}
		\hline
		& $\mathcal{X}_{1}$ & $\mathcal{X}_{2}$ & $\mathcal{X}_{3}$ & $\mathcal{X}_{4}$ & $\mathcal{X}_{5}$ & $Z_{1}$ & $Z_{2}$ & $Z_{3}$ & $Z_{4}$ & $Z_{5}$ \\ 
		\hline
		1 & 0.040 & 0.029 & 0.000 & 0.000 & 0.000 & 0.177 & 0.108 & 0.000 & 0.000 & 0.000 \\ 
		2 & 0.122 & 0.049 & 0.012 & 0.000 & 0.000 &  & 0.141 & 0.000 & 0.000 & 0.000 \\ 
		3 & 0.220 & 0.062 & 0.000 & 0.000 & 0.000 &  &  & 0.000 & 0.000 & 0.000 \\ 
		4 &  & 0.103 & 0.014 & 0.000 & 0.000 &  &  & 0.000 & 0.000 & 0.000 \\ 
		5 &  &  & 0.000 & 0.000 & 0.000 &  &  & 0.000 & 0.000 & 0.000 \\ 
		\hline
	\end{tabular}
\end{table}

The following example is related with a classification problem and so, the response is now binomial. We have included also an important number of needless variates to check how the algorithm deals with such situations. Consider $50$ iid variates distributed as $U[-1,1]$ and named, respectively, $\{X_1,X_2,Z_1,\ldots,Z_{48}\}$. The change in the names is to emphasize that only the first two variates are used in the true classification rule. Considering only the data points $X_1$ and $X_2$ belonging to the circular crown with $r=0.6$ and $R=1$ and its norm in $\R^2$,  $\norm{(X_1,X_2)}_2=\sqrt{X_1^2+X_2^2}$, we conform the first group with all data points that $0.6\le\norm{(X_1,X_2)}_2\le 0.8$ and the second fulfilling a similar condition $\norm{(X_1,X_2)}_2 >0.8$ as shown in Figure~\ref{fig:class} that shows a pairs plot of the first four variates $\{X_1,X_2,Z_1,Z_2\}$ for a simple run with sample size $N=1000$. Here the red and blue colors indicate the group for every datum. The choice of the best variates to construct the classification rule is an impossible task for a classical correlation coefficient due to the nonlinear nature of the problem.

\begin{knitrout}\footnotesize
	\definecolor{shadecolor}{rgb}{0.969, 0.969, 0.969}\color{fgcolor}\begin{figure}[htb!]
		{\centering \includegraphics[width=0.80\linewidth]{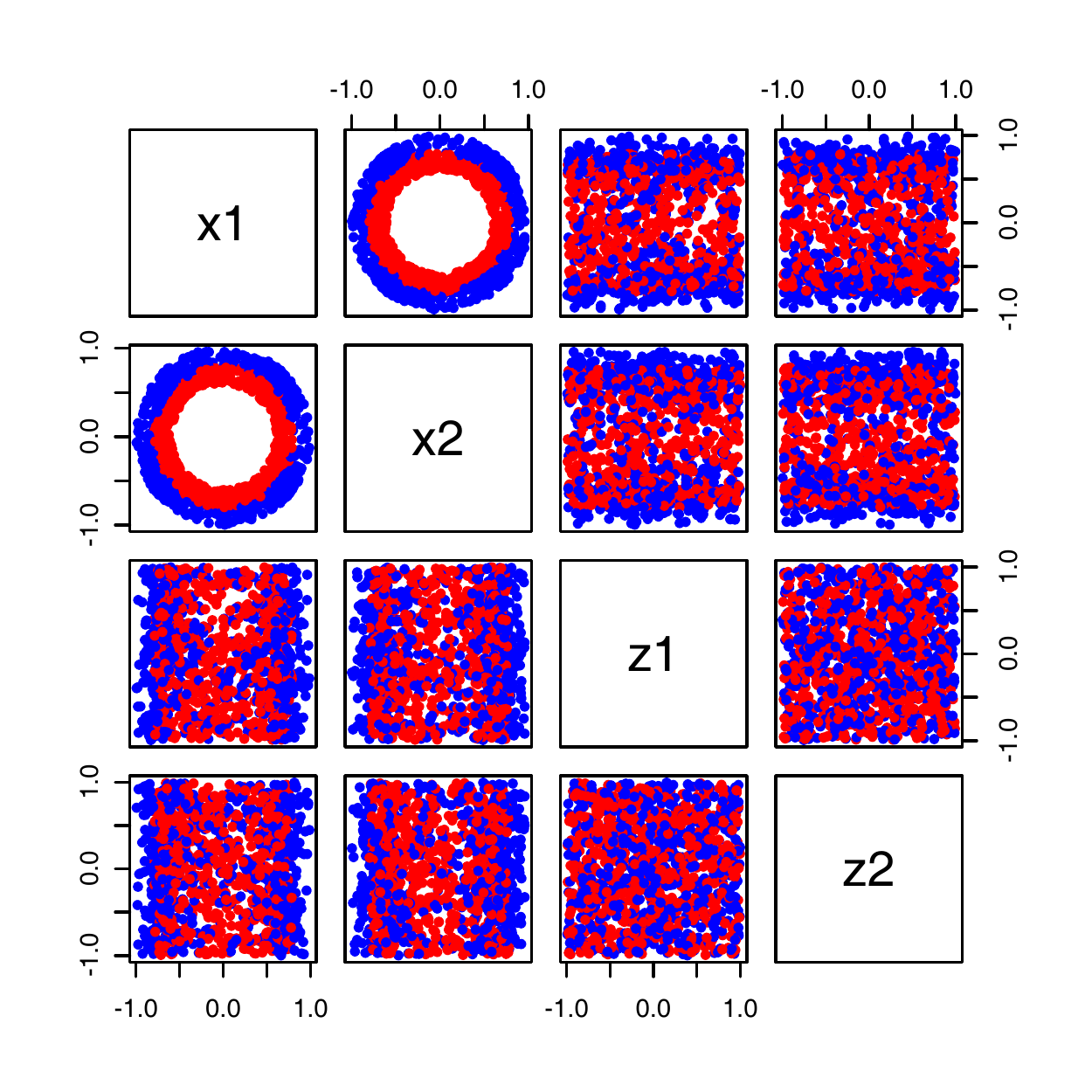} 			
		}		
		\caption[Example of four covariates for one run in the classification scenario]{Example of four covariates for one run in the classification scenario.}\label{fig:class}
	\end{figure}	
\end{knitrout}

We have applied the same algorithm described before to this case but using a generalized additive model: $\mathbb{E}\lrp{Y}=g^{-1}\lrp{\alpha+\sum_{j=1}^J f_j\lrp{X^{(j)}}}$ where $g$ is the link function. Note that the ``true'' model is not additive but a joint function of $X_1$ and $X_2$. The only change respect to the regression example is the way of computing the distance among elements of the categorical response. In this case, the distance matrix among elements of $Y$ only have two values: $0$ when both points belongs to the same group and $\sqrt{2}$ (the distance between $(1,0)$ and $(0,1)$) on the contrary case. We repeated this scenario $100$ times and applied the obtained model at each repetition to $200$ new points to check its prediction ability. The relevant variates $\{X_1,X_2\}$ for the model were selected in all the runs as the two first choices. The order is important because it is an indicator of the strength of the information provided by the variates. On the other hand, the proportion of times that the variates $\{Z_1,\ldots,Z_{48}\}$ enter the model were in the interval $\left[ 0, 0.07\right]$. The mean of misclassification error for new data was also excellent: $0.7$\%.


\section{Real data application}
\label{sec:appl}

We have applied our proposal to the Market Demand Energy in the Iberian Energy Market and its Price using the available information from different sources. Indeed, it is possible to create new variables as transformations/functions of the original ones. For instance, all the information included as a functional variable was also included as scalar covariates (each discretization point was a new one). We employed the FSAM/FAM model in all cases given its flexibility and availability not only for multivariate variables, but also for functional ones (see, for instance, \cite{Muller2008} or \cite{Febrero2013}).

\subsection{Energy Market Demand}
We have collected information about $N=2,459$ days from August, 2008 to December, 2014 of the Market Demand Energy at day $t$ and hour $H$ ($\textrm{En}_t(H)$) in the IntraDay Electricity Iberian Market. Our aim is to forecast this variable with the available information up to day $t-1$ obtained from the sources that measure energy or price. The meteorological information is considered concurrent in time because the forecasting of these variates can be done accurately. Here, the meteorological variables are taken from Madrid-Barajas airport as a coarse information. Detailed contributions of the meteorological information are expected that can occur when the energy demand is analyzed at low levels (substations, counties, etc) but no when we are dealing with national summaries. In any case, the station in Madrid seems to be a good choice due to two reasons that may have influence in high energy consumptions: Madrid has the highest population in Spain and it is surrounded by an important industrial belt. Finally, we consider also the calendar effects for every day although these variables may have redundant information with meteorological ones. For instance, temperatures and solar radiations are closely related with month as a calendar effect.

The following list summarizes all the variates involved: 
\begin{itemize}
	\item Energy Market Variables (source:www.omie.es): Daily profiles of Energy ($\textbf{En}$) and Price ($\textbf{Pr}$)  at $t-1$ and $t-7$.
	\item Total Load and Generated energy type (source:demanda.ree.es): Daily profiles of Load (every ten minutes) ($\textbf{Lo}$), Nuclear~($\textbf{Nu})$, Fuel/Gas~($\textbf{Fu}$), Coal~($\textbf{Ca}$), Combined Cycle~($\textbf{Cc}$), Solar~($\textbf{So}$), Wind Energy~($\textbf{WE}$), Hydroelectric~($\textbf{Hy}$), Cogeneration~($\textbf{Co}$), Rest~($\textbf{Re}$). 
	\item Meteorological information at Madrid-Barajas airport (source:aemet.es): Temperatures  ($T_{Max}$, $T_{Min}$, $T_{Med}$, $T_A=T_{Max}-T_{Min}$), Wind Speed ($WS$), Solar Radiation ($SR$), Precipitation, Pressure, $\ldots$.
	\item Every discretization value of the functional variates ($t-1$ and $t-7$).
	\item Categorical: Year($YY$), Month($MM$), Day-of-Week:\\ $DoW=\lrb{\Ind{Mon},\ldots,\Ind{Sun}}$. 
\end{itemize}

Specifically, we are interested on estimating the model \ref{eq:AM} where $Y_i=\textrm{En}_t(H)$, the number $J$ and the covariates $X_i^{(j)}$ must be selected from the above list but evaluated before day $t$ (except for meteorological and calendar variables), $f_j$ is the contribution of each variable and $\varepsilon_i$ is the error. The sample size is the number of days: $N=2,459$. 

Table~\ref{tab:energy}  summarizes the results of this application for four specific hours along the day where the variates are listed in order of inclusion into the model and the functional variables are marked in bold. In all those models, the deviance is mostly explained by the effect of the three first variables selected. As an example, the evolution in deviance of the model for $\textrm{En}_t(18)$ was 59.9\%, 71.9\%, 80.0\%, 80.7\% and 81.7\%. 
\begin{table}[h]
	\caption{Summary of models for energy with its selected variables (in order of entering)}
	\label{tab:energy}
	\centering
	\footnotesize
	\begin{tabular}{|c|l|c|c|}
		\hline
		Resp. & Covariates (in order) & Dev. expl. & $\sigma_\varepsilon$ \\ \hline\hline
		$\textrm{En}_t(6)$ & $\textrm{En}_{t-1}(6)$, $\textbf{En}_{t-7}$,  $WS$, $\textbf{Pr}_{t-1}$, $\textbf{Pr}_{t-7}$, $\textrm{En}_{t-1}(24)$, $\textbf{Re}_{t-1}$ & 84.4\% & 1207.6 \\ \hline
		$\textrm{En}_t(12)$ & $\textbf{En}_{t-7}$, $\textrm{En}_{t-1}(16)$, $DoW$, $\textbf{Cc}_{t-1}$, $\textrm{En}_{t-1}(23)$  & 84.2\% & 1540.3 \\ \hline
		$\textrm{En}_t(18)$ & $\textbf{En}_{t-7}$, $\textrm{En}_{t-1}(17)$, $DoW$, $WS$, $\textbf{Pr}_{t-1}$,  & 81.7\% & 1662.6 \\ \hline
		$\textrm{En}_t(24)$ & $\textrm{En}_{t-1}(24)$, $\textbf{Lo}_{t-7}$, $DoW$, $\textbf{Cc}_{t-1}$  & 86.0\% & 1221.0 \\ \hline
	\end{tabular}
	
\end{table}

In these models, the information of the first contributor tends to be the same as that of a week before rather than that  of the previous day except for the consumption in the early morning hours (06:00, 24:00). This suggests a strong weekly pattern in the market demand profile complemented with the appearance of the indicators for Day-of-Week as a contributor in the third place. The different behaviour of the early morning hours models comes from the fact that the energy demand at that hours is quite stable and has no main dependence on the activity of a given day of the week. Therefore, it is enough to consider the value of the day before. The inclusion into the models of variates like $\textbf{Pr}_{t-1}$ or $\textbf{Cc}_{t-1}$ although shocking at first glance, may be have a simple explanation. These two variates $\textbf{Pr}_{t-1}$ and $\textbf{Cc}_{t-1}$ are closely related to each other because, following the rules of Iberian Energy Market, the final price is fixed as the maximum of all types of energy required to cover the demand being the combined cycle, one of the expensive ones. The price is therefore higher when the proportion of energy produced by combined cycle power stations is also high. These variates  are probably included in the models to improve the prediction on the days characterized as having an energy demand that cannot be solved by using renewable sources (or other cheaper ones) although, in any case, its contribution is marginal.

\subsection{Energy Price}
The second application example corresponds to the negotiated price at day $t$ and hour $H$ ($\textrm{Pr}_t(H)$) in the Iberian Energy Market. The set of possible covariates is the same as in the previous example and the results are shown in Table~\ref{tab:price}. 

\begin{table}[h]
	\centering
	\caption{Summary of models for price with the selected variables (in order of entering)}
	\label{tab:price}
	\footnotesize
	\begin{tabular}{|c|l|c|c|}
		\hline
		Resp. & Covariates (in order) & Dev. expl. & $\sigma_\varepsilon$ \\ \hline\hline
		$\textrm{Pr}_t(6)$ & $\textbf{Pr}_{t-1}$, $T_A$, $T_{Max}$, $T_{Med}$, $SR$, $\textrm{Pr}_{t-1}(24)$, $\textbf{Cc}_{t-1}$ & 79.7\% & 6.93 \\ \hline
		$\textrm{Pr}_t(12)$ & $\textrm{Pr}_{t-1}(23)$, $T_{Max}$, $\textbf{Lo}_{t-7}$,  $\textrm{Pr}_{t-1}(9)$, $WS$,  $\textbf{Cc}_{t-1}$,  $\textbf{Pr}_{t-7}$, $\textbf{Cc}_{t-7}$ & 83.6\% & 6.46 \\ \hline
		$\textrm{Pr}_t(18)$ & $\textrm{Pr}_{t-1}(18)$, $DoW$, $WS$, $\textbf{Pr}_{t-7}$, $T_{Med}$, $\textrm{Pr}_{t-1}(16)$, $\textbf{Cc}_{t-1}$, $\textbf{Cc}_{t-7}$, $\textbf{Pr}_{t-1}$, $\textbf{WE}_{t-1}$ & 82.7\% & 6.85 \\ \hline
		$\textrm{Pr}_t(24)$ & $\textrm{Pr}_{t-1}(24)$, $\textbf{Pr}_{t-7}$, $\textbf{Cc}_{t-1}$, $\textbf{Cc}_{t-7}$, $YY$, $T_A$, $T_{Med}$ & 74.1\% & 7.05\\ \hline
	\end{tabular}
\end{table}

The main contributor in all price models is the price of the preceding day, which indicates a strong persistence of this variable. Also, all models include the combined cycle generation (in positions more or less advanced). Surprisingly now, the variables related with demand do not become part of the models. This is contrary to what is expected by the classical economic theory. The meteorological variates supply this gap. For instance, the model for $\textrm{En}_{t}(18)$ includes the mean of wind speed and the mean of temperature as covariates. These two variates can jointly explain the high price of energy on summer days with high temperatures but no wind, when the energy system must provide high amounts of energy (for feeding the air conditioning equipments) with low probability of using wind sources. The inclusion of other meteorological variates into the models follows similar rules. As commented before, it is not expected a high influence from meteorological variables because the information comes from a particular site (Barajas airport) and it would be hard to explain the price or the consumption of energy for a whole country using only the information of one site even though this is an important location. Surely, meteorological information is useful for predicting demand in small regions but its contribution for price can only be explained in terms of general rules. The calendar effects only appear in two models. The Day-of-Week effect appears in a prominent position in the model for $\textrm{En}_{t}(18)$ telling that difference in price between Mondays and Sundays (the maximum difference) is about twelve euros (the biggest differences are among weekend and non-weekend days). The Year effect only appears in the model for $\textrm{En}_{t}(24)$ showing a clearly increasing pattern along the years, which was particularly intense in the period 2009-2011. 

An important remark concerns the practical issues of the implementation of the distance correlation when the sample size grows. The main difficulty is unrelated to the ease of the implementation rather with the memory consumption of the method. The best strategy for implementing this procedure making an extensive use of the distance correlation with respect to the same covariates, is to compute and store the distance matrix for every covariate in advance. However, strategy is impossible when the sample size or the number of covariates grows. The overall consumption of memory when storing the distance matrices for $p$ covariates with $N$ elements each is $(p+1)\lrp{N(N-1)/2-N}$. In this example, we have $p=300$ covariates (many of which correspond to the discretization points of the functional ones) with a sample size $N=2459$. This amount typically exceeds the available resources of a desktop computer (even for the powerful ones). To overcome this difficulty, the obvious response is to use a HPC facility where the computation of distance correlation for each pair can be distributed along the available nodes (storing the distance matrices in the shared storage device). When it is impossible to access to a HPC facility, one may still compute the distance correlation for an arbitrary sample size by simply dividing all the computations in blocks using submatrices of dimension $L\times L$. As an example, Table~\ref{tab:ram} shows the maximum consumed memory and the execution time (in seconds) when the test of independence based on distance correlation for different sample sizes and two different dimensions of the submatrices is executed in a Intel Core i7-3770 with 32GB of RAM. The last row of the table shows that with $N=25000$ the computer was unable to reserve enough memory. The submatrix trick can manage that sample size maintaining a limit for the memory consumption, but it does so with the cost of larger execution times. The execution time using $L=500$ seems slightly better than with $L=1000$. However we cannot conclude a general rule. For instance, for $N=10000$ the execution times with $L=100,250$ and $2000$ were, respectively, $52.10, 39.84$ and $64.93$. 

\begin{table}
	\centering
	\caption{Maximum memory consumption and computation times (s.) for computing the test of independence using distance correlation with different computational strategies involving submatrices.}
	\label{tab:ram}
	\footnotesize
	\begin{tabular}{|c|r|r|r||r|r|r|}
		\hline
		& \multicolumn{3}{c||}{Memory consumption} & \multicolumn{3}{|c|}{Time(sec.)} \\ \cline{2-4}\cline{5-7}
		N & Direct & $L=500$ & $L=1000$ & Direct & $L=500$ & $L=1000$ \\ \hline\hline
		1000 & 7.63MB  & 7.63MB & 7.63MB & 0.32 & 0.94 & 0.48 \\ \hline
		2500 & 47.68MB  & 7.63MB & 30.52MB & 1.58 & 4.40 & 5.40 \\ \hline
		5000 & 190.7MB  & 7.63MB & 30.52MB & 6.57 & 13.08 & 15.47 \\ \hline
		10000 & 762.9MB  & 7.63MB & 30.52MB & 26.83 & 45.91 & 53.00 \\ \hline
		25000 & CRASH  & 7.63MB & 30.52MB & -- & 290.13 & 309.91 \\ \hline
		
	\end{tabular}
\end{table}

\section{Conclusions}
In this paper, we have introduced an algorithm that automatically selects the variates for a regression or classification model. The procedure operates in a forward way adding a variable to the model at each iteration. The key of the whole procedure is the extensive use of the distance correlation that presents two important advantages: the choice of the variate is made without considering a model and it is possible to compute this quantity for variates of different nature (functional, multivariate, circular, ...) as it is only computed from distances. The simplicity of the latter is also its main drawback because the number of operations (and memory consumption) is of a quadratic order respect to the sample size. But fortunately, it is possible to compute the distance correlation when the sample size is huge while the consumption of resources is maintained under certain limitations. 
Our proposal is presented is a very general way although in the applications in this paper, we have restricted ourselves to additive models that offer a balanced compromise between predictive ability and simplicity. The obtained results are quite promising in scenarios where no competitors are available because no other procedure can deal with variates of different nature in a homogeneous way. As a final comment, the procedure was applied to a real problem related with the Iberian Energy Market (Price and Demand) where the number of possible covariates is really big. The algorithm was able to find synthetic regression models offering interesting insights about the relationship among the response and the covariates. The final selected models mix functional, scalar and categorical information.  


\section*{Acknowledgement}
	The authors acknowledge financial support from Ministerio de Econom\'ia y Competitividad grant MTM2016-76969-P and European Regional Development Fund (ERDF). The authors are also grateful to the Guest Editor, Associate Editor and two anonymous referees for their helpful and valuable comments.

\bibliographystyle{abbrvnat}


\end{document}